\documentclass[twocolumn,a4paper,10pt]{article}
\usepackage{graphicx}

\title{Combined holographic-mechanical optical tweezers:\\Construction, optimisation and calibration}
\author{Richard D.L. Hanes, Matthew C. Jenkins, Stefan U. Egelhaaf\\Condensed Matter Physics Laboratory,\\Heinrich-Heine University, 40225 D\"usseldorf, Germany}
\date{\today}

\begin{document}

\maketitle

\begin{abstract}
A spatial light modulator (SLM) and a pair of galvanometer-mounted mirrors (GMM) were combined into an optical tweezers set-up. This provides great flexibility as the SLM creates an array of traps which can be moved smoothly and quickly with the GMM. To optimise performance, the effect of the incidence angle on the SLM with respect to phase and intensity response was investigated. Although it is common to use the SLM at an incidence angle of 45 degrees, smaller angles give a full $2\pi$ phase shift and an output intensity which is less dependent on the magnitude of the phase shift. The traps were calibrated using an active oscillatory technique and a passive probability distribution method. 
\end{abstract}

\section{Introduction}
Since the invention of optical tweezers, they have had a major impact on the way nano-sized objects can be manipulated, especially in biology~\cite{Svoboda1994, Sheetz1998, Greulich1999} and colloidal physics~\cite{Molloy2002, Ashkin1997}. The first realisation was shown by Ashkin in 1970~\cite{Ashkin1970} where he demonstrated that spheres in a solvent of lower refractive index are subject to two forces. First, the gradient force draws particles into the centre of the laser beam due to the radial gradient in the beam intensity. Second, the scattering force accelerates the particles in the direction of the beam due to radiation pressure. If the gradient force dominates the scattering force, which is achieved by tightly focusing the beam, a stable trap in three dimensions is created~\cite{Ashkin1986}, giving rise to optical tweezers~\cite{Molloy2002, Grier2003, Otto2008, Bartlett2002}. The trap can be moved within the sample by steering the laser beam using mirrors, which are typically mounted on galvanometers. Multiple-trap tweezers have been realised by splitting one beam into several traps in two or three dimensions using acousto-optic modulators~\cite{Simmons1996, Vossen2004, Biancaniello2006} or phase holograms~\cite{Dufresne1998, Reicherter1999, Liesener2000, Dufresne2001, Curtis2002, Sinclair2004a}. Both of these techniques allow for motion of the traps, with their specific strengths and limitations~\cite{Molloy2002, Neuman2004}. In addition to exerting a force on particles and thereby manipulating them, tweezers can also be used to measure forces in the pico- and femtonewton range~\cite{Gutsche2007, Ghislain1994, Rohrbach2005a}.

This paper discusses a specific implementation of optical tweezers which combines a spatial light modulator (SLM) with galvanometer-mounted mirrors (GMM). The SLM can generate complex arrays of traps in three dimensions, but the trap dynamics are limited by the refresh rate of the device. The GMM can translate the trap array in a fast and continuous way, increasing the flexibility compared to an SLM alone. Furthermore, by combining both devices, multiple traps can be calibrated simultaneously, allowing them to be used for force measurements. 

\section{Optical tweezers set-up}
\begin{figure}
\centering
\includegraphics[width=1.0\linewidth]{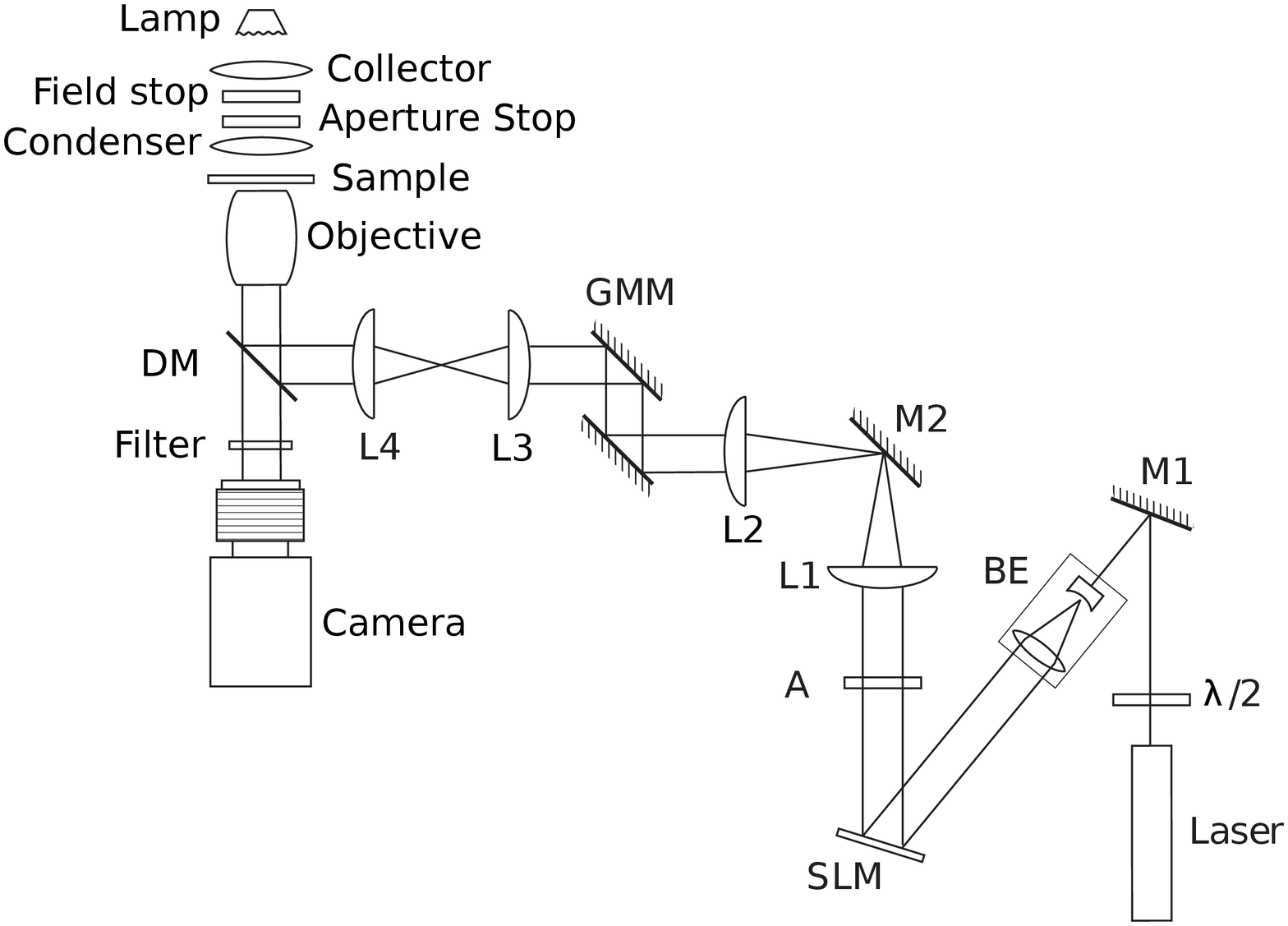}
\caption{\label{fig:setup} Schematic diagram of the optical tweezers set-up combining a pair of galvanometer-mounted mirrors (GMM) and a spatial light modulator (SLM). (Abbreviations are defined in the main text.)}
\end{figure}

A schematic representation of the set-up is shown in fig \ref{fig:setup}. A diode-pumped solid-state laser with $\lambda$ = 532~nm and a maximum continuous output power of 1.5~W (Laser Quantum Ventus 532-1500) is used. The polarisation angle is rotated by a half-wave plate ($\lambda$/2), necessary for the SLM later in the optical path. The beam is reflected from mirror M1 (all mirrors from Thorlabs), then passes through a 10x beam expander BE (Newport NT55-578), which increases the beam diameter from 1.4~mm to 14~mm, the length of the longest edge of the SLM. It strikes the SLM (Holoeye LCR-2500) at an angle of incidence $\alpha$ = 22.5$^{\circ}$ and passes through the analyser (A) which, together with the half-wave plate ($\lambda$/2), is necessary to optimise the device performance (section~\ref{sec:SLM}). The light reflected by the SLM is then imaged onto the GMM  by a telescope formed by lenses L1 ($f_{1}$=175~mm planoconvex, all lenses from Edmund Optics) and L2 ($f_{2}$=100~mm planoconvex) which also reduces the diameter from 14~mm to $d_{\mathrm{GMM}}$ = 8~mm. Between the two lenses the light is reflected vertically by mirror M2, to align the beam with the microscope epi-flourescence port. The GMM (Nutfield Technology Inc. Quantum Scan 30) consist of two galvanometer-mounted mirrors with perpendicular axes of rotation for two-dimensional in-plane translation of the traps. From here, the beam traverses the second telescope (fig~\ref{fig:galvoangle}) formed by lenses L3 ($f_{3}$=125~mm planoconvex) and L4 ($f_{4}$=100~mm planoconvex) which reduces the beam to its final diameter of $d_{\mathrm{Obj}}$ = $(f_{\mathrm{4}}/f_{\mathrm{3}})~d_{\mathrm{GMM}}$ = $(\theta_{\mathrm{GMM}}/\theta_{\mathrm{Obj}})~d_{\mathrm{GMM}}$ = 6.4~mm, to slightly overfill the back aperture of the microscope objective. Combined imaging and trapping is facilitated by a dichroic mirror DM (transmission in the range 525-540~nm is lower than 12~\%, Chroma Technology Corp. z532dcrb) which reflects the laser light into a 1.4 NA 100x Nikon objective. For imaging we use an inverted microscope (Nikon TE2000-U). The intense laser light is separated from the imaging light by a filter (notch filter, transmission in the range 525-538~nm is lower than 10$^{-4}$~\%, Chroma Technology Corp. z532nf). The filter, the dichroic mirror DM and the lens L4 are housed within a home-built filter cube inside the microscope. The other optical components are mounted directly onto the optical table on which the microscope is also fixed. For capturing images a CMOS camera (Pixelink PL-B742F) controlled by LabView (National Instruments) is used.

\section{Galvanometers}
\begin{figure}
\includegraphics[width=1.0\linewidth]{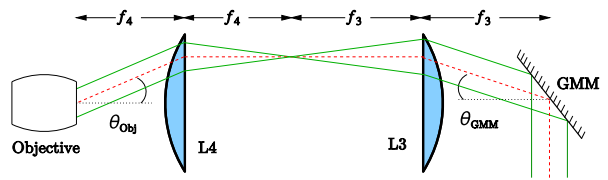}
\caption{\label{fig:galvoangle} (Colour online) The rotation angle $\theta_{\mathrm{GMM}}$ of the galvanometer-mounted mirrors (GMM) results in a tilt $\theta_{\mathrm{Obj}}$ at the back aperture of the objective. It is constrained by the geometry of the setup. Note that the GMM consists of two mirrors; only one is shown for clarity.}
\end{figure}

The galvanometer-mounted mirrors (GMM) are used to steer the beam into lens L3, part of the final telescope in the system (fig~\ref{fig:galvoangle}). The telescope images the mirrors onto the back aperture of the objective, so that a rotation of the mirrors through an angle $\theta_{\mathrm{GMM}}$ corresponds to a rotation of the beam at the aperture through an angle $\theta_{\mathrm{Obj}}$, and an in-plane translation of the trap(s). Ideally both GMM would lie and rotate about a single point, but in practice this is not possible and they are thus slightly separated (by 2.4~cm). This leads to a movement of the beam across their surfaces when they rotate and, as a result, the laser intensity through the objective varies. However, this variation is less than 5\% across the whole field of view, and all trap strength measurements were performed at the same position within the field of view.

The diameter and focal length of lens L3 give rise to a maximum useful mirror rotation angle $\theta^{\mathrm{max}}_{\mathrm{GMM}}$ and hence a maximum translation distance within the sample. The diameter of lens L3 is 25.4~mm and the beam diameter at this point is $d_{\mathrm{GMM}}$ = 8~mm, giving a maximum displacement of the beam centre from the centre of lens L3 of (25.4~mm - 8~mm)/2 = 8.7~mm. Furthermore, the distance between the central point between the two GMM and lens L3 is $f_{3}$=125~mm resulting in a maximum rotation angle $\theta^{\mathrm{max}}_{\mathrm{GMM}} = \tan^{-1}(8.7$~mm$/125$~mm$) = 4.0^{\circ}$. At the objective this corresponds to a maximum rotation $\theta^{\mathrm{max}}_{\mathrm{Obj}} = (f_{\mathrm{4}}/f_{\mathrm{3}})~\theta^{\mathrm{max}}_{\mathrm{GMM}} = 5.0\,^{\circ}$ and an in-plane displacement of $\pm83$~$\mathrm{\mu}$m. This displacement was determined experimentally and calibrated as follows: the filter was removed from the imaging path and the lowest available laser power used. This allowed us to observe the reflection of the beam from the interface between the cover slip and the solvent. The reflection was quantitatively tracked using a custom-written LabView routine (section~\ref{sec:tracking}). By moving a trap in the $x$ and $y$ directions the GMM voltage could be related to the lateral displacement in pixels (147 pixels/volt). No significant differences were found for different positions or directions within the field of view. The pixels were then related to absolute measures by comparison to the image of a Richardson test slide, which gave a value of 68.5~nm/pixel and hence a maximum displacement of $\pm83$~$\mathrm{\mu}$m.

The GMM are controlled by a digital-to-analogue card DA (National Instruments PCI-6014) and home-written LabView software. The output range from the DA card is -10~V to +10~V in steps of 5.8~mV,  corresponding to a maximum rotation range of the mirrors of $\pm48^{\circ}$ with increment $0.03^{\circ}$. Since we are using a smaller range of angles $|\theta_{\mathrm{GMM}}| {<} \theta^{\mathrm{max}}_{\mathrm{GMM}} {=} 4.0^{\circ}$, we incorporated a 10x voltage divider between the computer and galvanometer control electronics (Quantum Drive 3000) resulting in a maximum possible rotation of $\pm4.8^{\circ}{>}\theta^{\mathrm{max}}_{\mathrm{GMM}}$ with increment $0.003^{\circ}$. Details of trap dynamics are discussed below (section~\ref{sec:time}). 

\section{SLM characterisation\label{sec:SLM}}
To optimise the performance of the tweezers system the SLM was characterised in detail. Optimum performance entails a high relative intensity of the wanted first order projections with respect to the unwanted zero and higher order intensities, resulting in the desired size and shape of each trap without the presence of ghost traps. An SLM working at optimum efficiency (ignoring the algorithm used to calculate the phase holograms~\cite{DiLeonardo2007}) will be able to modify the phase of the incoming light by 2$\pi$, without significantly affecting its amplitude. Our SLM uses nematic liquid crystals to control the refractive index at each pixel, making it highly sensitive to the incoming polarisation orientation. It also causes a slight depolarisation which reduces the efficiency, and requires an analyser to recover a well defined polarisation.

\begin{figure}
\centering
\includegraphics[height=16cm]{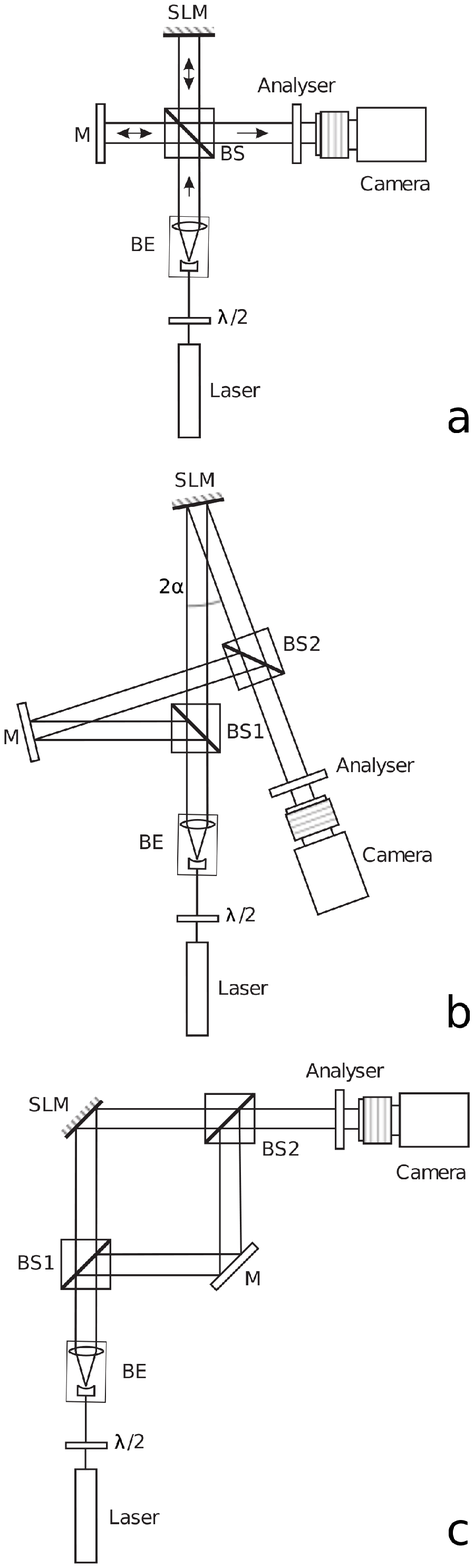}
\caption{\label{fig:charsetups}Interferometers used for determining the intensity and phase as a function of different incidence angles $\alpha$. (a) $\alpha=$ 0$^{\circ}$, (b) 3$^\circ<\alpha\leq$ 22.5$^{\circ}$ and (c) $\alpha$ = 45$^{\circ}$. For the intensity measurements, the beam directed towards the mirror M is covered and the camera used as a power meter. (Abbreviations are defined in the main text.)}
\end{figure}

A Holoeye LCR-2500 SLM is used which gives a 2$\pi$ phase shift from 400~nm to 700~nm when illuminated at a small angle of incidence (manufacturer's specifications). For this device the effects of changing the incidence angle $\alpha$ on the output intensity and phase were investigated experimentally. The different $\alpha$ require interferometers with different geometries (fig~\ref{fig:charsetups}). The set-up used for zero angle incidence geometry represents a Michelson-Morley interferometer~\cite{Michelson1887} (fig~\ref{fig:charsetups}a). The beam is first expanded by a beam expander BE and then split by a beam splitter cube BS (B.Halle TWK-20) between a mirror M and the SLM. The reflections are collected at the same BS, passed through an analyser and recorded on the camera. The incoming beam polarisation is rotated using a half wave plate ($\lambda$/2). For incidence angles 3$^\circ<\alpha\leq$ 22.5$^{\circ}$ we used another interferometer (fig~\ref{fig:charsetups}b). This set-up contains two beam splitter cubes, BS1 to divide the incoming beam and BS2 to recombine them. A symmetric set-up with nearly equal path lengths is required due to the limited coherence length of our laser (about 1~cm). A similar set-up, but with rectangular geometry, a Mach-Zehnder interferometer~\cite{Zehnder1891, Mach1892}, is applied for an incidence angle $\alpha=$ 45$^{\circ}$ (fig~\ref{fig:charsetups}c). 

\subsection{Output intensity}
\begin{figure}
\centering
\includegraphics[width=1.0\linewidth]{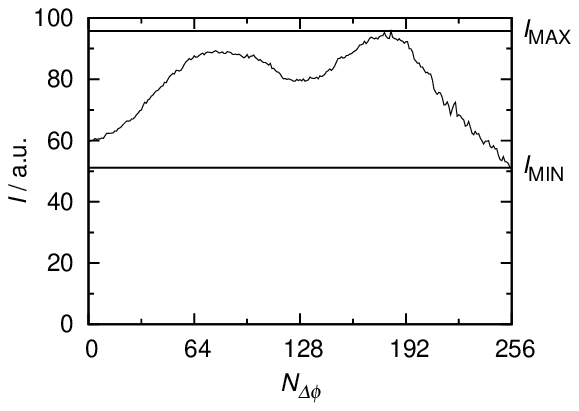}
\caption{\label{result:intensitycurve}Output intensity $I$ as a function of phase shift $N_{\Delta\phi}$ introduced by the SLM for an incidence angle $\alpha$ = 22.5$^{\circ}$ and input and output polarisation $\varphi_{\mathrm{i}}$=140$^{\circ}$ and $\varphi_{\mathrm{o}}$=0$^{\circ}$ respectively.}
\end{figure}

Ideally, the SLM would introduce a phase shift without affecting the output intensity. Since this is in general not achieved for all conditions, the SLM was characterised to find its optimum configuration. For different incidence angles $\alpha$, we determined the output intensity $I$ as a function of phase shift at the SLM $\Delta\phi$ for different polarisations of the input $\varphi_{\mathrm{i}}$ and output $\varphi_{\mathrm{o}}$ beams. The mirror arm of the set-up was blocked using a beam dump, and the intensity $I$ reflected from the SLM recorded with the camera. The intensity recorded across all pixels of the camera was averaged to improve statistics. It was ensured that when the laser was off the reading of the camera was zero and that saturation of the camera was avoided. The input, $\varphi_{\mathrm{i}}$, and output, $\varphi_{\mathrm{o}}$, polarisations were selected using the half-wave plate ($\lambda$/2) and analyser. The polarisations were defined relative to the laboratory vertical and increased clockwise in the direction of beam propagation. The phase shift $\Delta\phi$ introduced by the SLM was then increased in steps and the intensity $I$ measured. An increment of 1 out of the 256 levels that our SLM offers, i.e. 0 $\le N_{\Delta\phi}<$ 256, was used. An example of the dependence of the intensity on the phase shift is shown in fig~\ref{result:intensitycurve}. 

\begin{figure}
\centering
\includegraphics[width=0.55\linewidth]{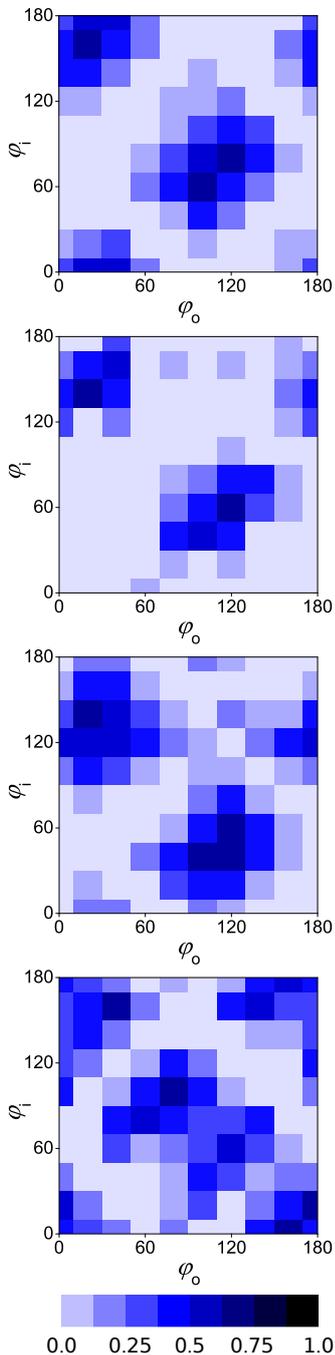}
\caption{(Colour online)\label{result:intensity} Ratio between minimum and maximum intensity $I_{\mathrm{MIN}}$/$I_{\mathrm{MAX}}$ (fig \ref{result:intensitycurve}) as a function of input, $\varphi_{\mathrm{i}}$, and output, $\varphi_{\mathrm{o}}$, polarisation for different incident angles $\alpha=$ 0$^{\circ}$, 4$^{\circ}$, 22.5$^{\circ}$ and 45$^{\circ}$ (top to bottom). Plateau regions are observed at ($\varphi_{\mathrm{i}}$, $\varphi_{\mathrm{o}}$) $\approx$ (160, 20) and (60, 100).}
\end{figure}

As a measure of the flatness of the response, we use the ratio between minimum and maximum intensity $I_{\mathrm{MIN}}$/$I_{\mathrm{MAX}}$, therefore values close to one indicate an output intensity $I$ almost independent of the phase shift $\Delta\phi$. (Note that sometimes the output amplitude $A$ is considered with $I_{\mathrm{MIN}}$/$I_{\mathrm{MAX}}$ = ($A_{\mathrm{MIN}}$/$A_{\mathrm{MAX}}$)$^{2}$~\cite{Martin-Badosa1997}.) The dependence of $I_{\mathrm{MIN}}$/$I_{\mathrm{MAX}}$ on $\varphi_{\mathrm{i}}$ and $\varphi_{\mathrm{o}}$ is shown in fig~\ref{result:intensity} for different incidence angles $\alpha$. We covered a range 0$^{\circ}\le\varphi_{\mathrm{i,o}}\le$180$^{\circ}$ in increments of 20$^{\circ}$ with the results at $\varphi_{\mathrm{i,o}}=0^{\circ}$ and $\varphi_{\mathrm{i,o}}=180^{\circ}$ being identical within experimental uncertainties. 

The intensity maps (fig~\ref{result:intensity}) show that there are clear plateaux with $I_{\mathrm{MIN}}$/$I_{\mathrm{MAX}}$ close to 1 for all values of $\alpha$. However, a tweezer set-up with $\alpha=0^\circ$ involves passing a BS twice which reduces the intensity transmitted to the objective to a quarter and $\alpha=4^{\circ}$ requires long beam paths increasing the sensitivity to vibrations. Both configurations were therefore rejected and only $\alpha= $22.5$^{\circ}$ and 45$^{\circ}$ were investigated further. In addition to hardly affecting the output intensity (implying $I_{\mathrm{MIN}}$/$I_{\mathrm{MAX}} \approx$ 1), the SLM should also provide $\Delta\phi\ge2\pi$, with $\Delta\phi$ not significantly larger than $2\pi$ to maintain a large range of $N_{\Delta\phi}$. This is discussed in the following section.

\subsection{Phase change}\label{sec:phase}
\begin{figure}
\centering
\includegraphics[width=1.0\linewidth]{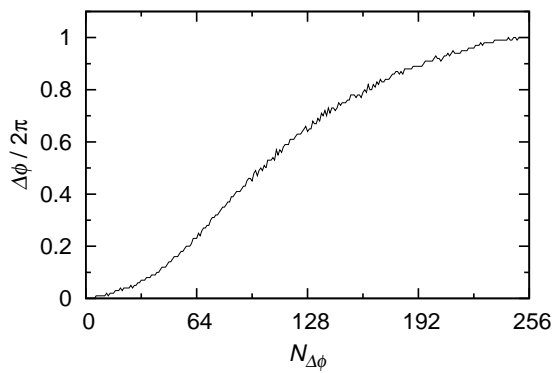}
\caption{\label{result:phasecurve}Actual phase shift $\Delta\phi$ as a function of $N_{\Delta\phi}$ for an incident angle $\alpha$ = 22.5$^{\circ}$ and input and output polarisation $\varphi_{\mathrm{i}}$=140$^{\circ}$ and $\varphi_{\mathrm{o}}$=0$^{\circ}$ respectively.}
\end{figure}

We now relate $N_{\Delta\phi}$ to the actual phase shift $\Delta\phi$ (fig \ref{result:phasecurve}) and determine the maximum phase shift accessible for the different parameters; $\varphi_{\mathrm{i}}$, $\varphi_{\mathrm{o}}$ and $\alpha$. The same set-ups are used but with the beam dump removed from the mirror arm of the interferometer (fig \ref{fig:charsetups}). The mirror M was aligned to produce vertical fringes with a fringe width of 64 pixels, which was shown to provide optimum conditions considering processing errors and statistics due to the number of detectable fringes on our camera (1280 x 1024 pixels)~\cite{Martin-Badosa1997}. The two halves (top and bottom) of the SLM are programmed separately, one being held at a constant $N_{\Delta\phi}$=0, the other increased in steps of 1. As the difference in $N_{\Delta\phi}$ and hence the optical path length between the two halves of the SLM increases, the fringes detected on the two halves of the camera shift relative to each other. To recover the phase shift $\Delta\phi$, we took cross-sections from the top and bottom of the image. By cross-correlating these two signals, and finding the first maximum, we recover the shift in pixels. This is converted to actual phase shift $\Delta\phi$ by multiplication by 2$\pi$ and division by the period of the fringes in pixels, found from the Fourier transform of one of the signals.

\begin{figure}
\centering
\includegraphics[height=15cm]{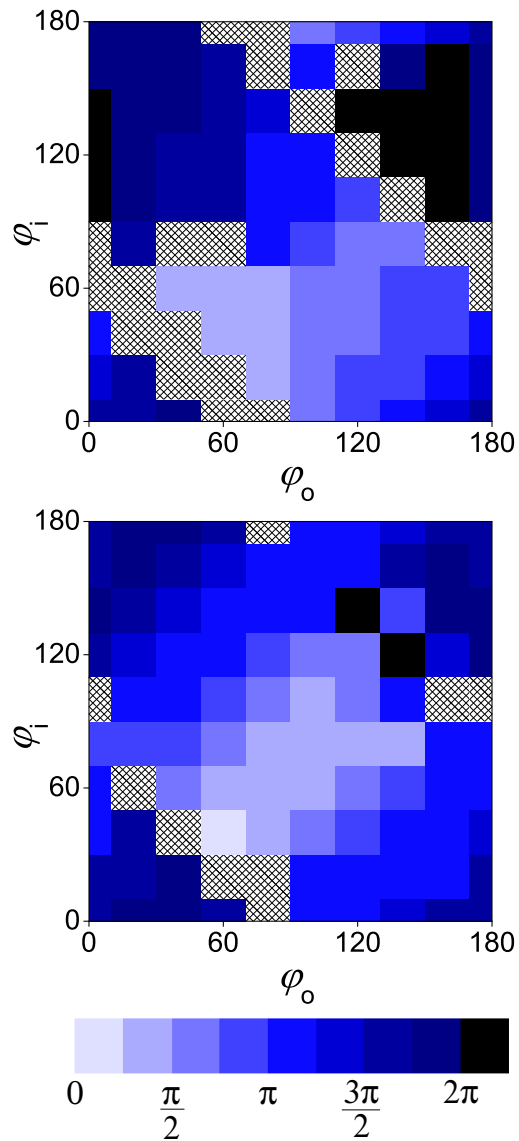}
\caption{(Colour online.) \label{fig:phasemap}Maximum range of the actual phase shift $\Delta\phi$ as a function of input, $\varphi_{\mathrm{i}}$, and output, $\varphi_{\mathrm{o}}$, polarisation for different incident angles $\alpha=$ 22.5$^{\circ}$ and 45$^{\circ}$ (top, bottom). The cross-hatched areas represent configurations where the intensity of one of the fringe signals was too small to be measured, and therefore $\Delta\phi$ could not be determined.}
\end{figure}

We investigated the maximum accessible range of the phase shift for incident angles $\alpha=22.5^\circ$ and $\alpha=45^\circ$ for the same range of input, $\varphi_{\mathrm{i}}$, and output, $\varphi_{\mathrm{o}}$, polarisations as above, i.e.~$0^\circ \le \varphi_{\mathrm{i,o}} \le 180^\circ$ in increments of 20$^\circ$ (fig~\ref{fig:phasemap}). Close to the optimum conditions, we used increments of 1$^\circ$ (data not shown). We found that, for $\alpha=45^{\circ}$, there was no configuration which gave a full $2\pi$ phase shift with $I_{\mathrm{MIN}}/I_{\mathrm{MAX}} > 0.1$. Therefore we use the $\alpha=22.5^\circ$ configuration which gives optimum results for $\varphi_{i}=140^\circ$ and $\varphi_{o}=0^\circ$. These parameters result in a phase shift $\Delta\phi=2.0\pi$ and an intensity ratio $I_{\mathrm{MIN}}/I_{\mathrm{MAX}}=0.54$. The actual phase shift $\Delta\phi$ as a function of $N_{\Delta\phi}$ for this configuration is shown in figure \ref{result:phasecurve}. The non-linearity is corrected for by the software.

\section{Trap calibration and use}
\subsection{Active method: Oscillatory drag force}\label{sec:tracking}
\begin{figure}
\centering
\includegraphics[width=1.0\linewidth]{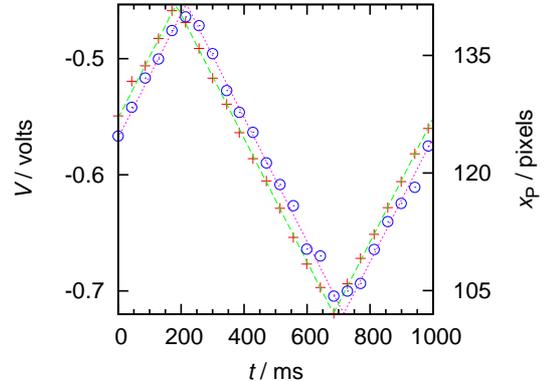}
\caption{(Colour online.)\label{result:triangular} Voltage V sent to galvanometer (+) and particle position $x_{\mathrm{P}}$ ($\circ$) together with fits as a function of time $t$. Polystyrene-sulfonate particles with a radius $R =$ 2.0~$\mathrm{\mu}$m in water at a distance to the coverslip of about 15~$\mathrm{\mu}$m were subjected to an oscillatrory optical tweezer with a laser power of 11.1~mW at a temperature of T=25 $\pm 1~^{\circ}$C. We determined a time lag between particle and trap position $\Delta t$ = 20.8 $\pm$ 0.8~ms giving a trap stiffness $\kappa$ = $(1.6 \pm 0.1) \times 10^{-3}$~pN/nm.}
\end{figure}

We use an oscillatory calibration technique, based on the drag force method~\cite{Kuo1993}. However, here, instead of the stage the SLM-generated trap is moved using the GMM, and a precise positional calibration of the stage is thus not required. A triangular wave signal with frequency $\nu$ = 1~Hz is sent to the galvanometers. The reflection of the laser from the cover glass-water interface was tracked using one of the pattern matching algorithms in LabView. It convolves the image of the laser spot with a template image of the reflected laser, to determine potential locations of the laser beam. Subsequently, the locations are refined to sub-pixel accuracy. The size of the tracked object (in pixels) determines the accuracy of the tracking as well as the maximum number of objects in the field of view. In our experiments the objects are typically 60 x 60 pixels, resulting in an accuracy of 15~nm, determined by tracking a reflected laser spot. With our camera (full frame 1280 x 1024 pixels at 27 frames/s, or e.g. 100 x 100 pixels at 500 frames/s) this results in a maximum of 360 objects at a time resolution of 37~ms, or one object at a time resolution of 2~ms. The position of the laser $x_{\mathrm{L}}(t)$ and the voltage sent to the GMM $V(t)$ were recorded and $x_{\mathrm{L}}(t)\propto \mathrm{sin^{-1}(cos(}2\pi\nu(t+\Delta t_{\mathrm{L}})))$ and $V(t)\propto \mathrm{sin^{-1}(cos(}2\pi\nu(t+\Delta t_{\mathrm{V}})))$ fitted to the data. The time delay $\Delta t_{0}=\Delta t_{\mathrm{L}}-\Delta t_{\mathrm{V}}$ provides the processing time of the hardware and software system; for a frame rate of 70 frames/s it is typically 10~ms. We subsequently trap a particle with radius $R$ and perform a similar procedure, but this time recording the particle position $x_{\mathrm{P}}(t)$ and using $x_{\mathrm{P}}(t)\propto \mathrm{sin^{-1}(cos(}2\pi\nu(t+\Delta t_{\mathrm{P}})))$ as a fit function to obtain the time delay $\Delta t_{1}=\Delta t_{\mathrm{P}}-\Delta t_{\mathrm{V}}$ (fig~\ref{result:triangular}). The time lag $\Delta t = \Delta t_{1}-\Delta t_{0}$ can be related to the trap stiffness $\kappa=6\pi\eta R/\Delta t$ by balancing the drag force due to particle velocity $v$ in a viscous medium (viscosity $\eta$) $F_{\eta}=6\pi\eta Rv$ and the trap force $F_{\mathrm{trap}}=\kappa\Delta x=\kappa v \Delta t$, assuming a harmonic potential. A typical example using polystyrene sulfonate particles in water with $R = 2.0 \pm 0.1$~$\mathrm{\mu m}$ (Interfacial Dynamics Corporation), a laser power of 11.1~mW, at temperature $T = 25 \pm 1$~$^{\circ}$C (leading to $\eta = (0.89 \pm 0.02) \times 10^{-3}~$Pa s) and a distance to the coverslip of about 15~$\mathrm{\mu m}$ is shown in figure~\ref{result:triangular}. The fit resulted in $\Delta t$ = 20.8 $\pm$ 0.8~ms, and thus $\kappa$ = $(1.6 \pm 0.1) \times 10^{-3}$~pN/nm and a maximum excursion of the particle from the trap center $\Delta x_{\mathrm{MAX}}$ = $134 \pm 5$~nm. The assumption that the trap potential is harmonic up to $\Delta x_{\mathrm{MAX}}$ was confirmed (section \ref{sec:passive} and fig~\ref{result:histogram}). 

The trap stiffness $\kappa = \xi/\Delta t$ depends on the friction coefficient $\xi = 6\pi\eta R$ which above was calculated from $\eta$ and $R$. It can, however, also be determined independantly from the free diffusion coefficient of the particle in the bulk $D=k_BT/\xi$ to give $\kappa = k_BT/D\Delta t$. $D$ was determined from the mean squared displacement in two dimensions $<r^2> = 4Dt$ by recording the particle motion while diffusing. In order to track the particle accurately, it was kept close to the imaging plane and 15~$\mathrm{\mu} m$ from the coverslip using the tweezers. The tweezers were programmed to trap and release the particle periodically with a frequency of 1~Hz which was realised by jumping the tweezers to a distant point using the GMM. To recover the free diffusion coefficient $D$, the data corresponding to the periods when the particle was influenced by the tweezers were discarded and the traces of the particle while it was freely diffusing shifted so that the data at the boundaries overlapped. To test this method we used the experimentally determined $D$ to calculate the radius $R = \mathrm{k_B} T / 6\pi\eta D$ of the same particles as above ($R = 2.0 \pm 0.1 \mathrm{\mu m}$ manufacturers specification's) making sure the selected particles are representative; very large or small particles were rejected based on their size in the recorded images. We obtain an average particle size of $2.04 \pm 0.06 \mathrm{\mu m}$ (which included the polydispersity of the sample) with an error bound of an individual particle radius of about 2\%. This compares very well with the manufacturer's specifications and indicates a successful determination of the friction coefficient $\xi$.

\subsection{Passive method: Probability distribution}\label{sec:passive}
\begin{figure}
\centering
\includegraphics[width=1.0\linewidth]{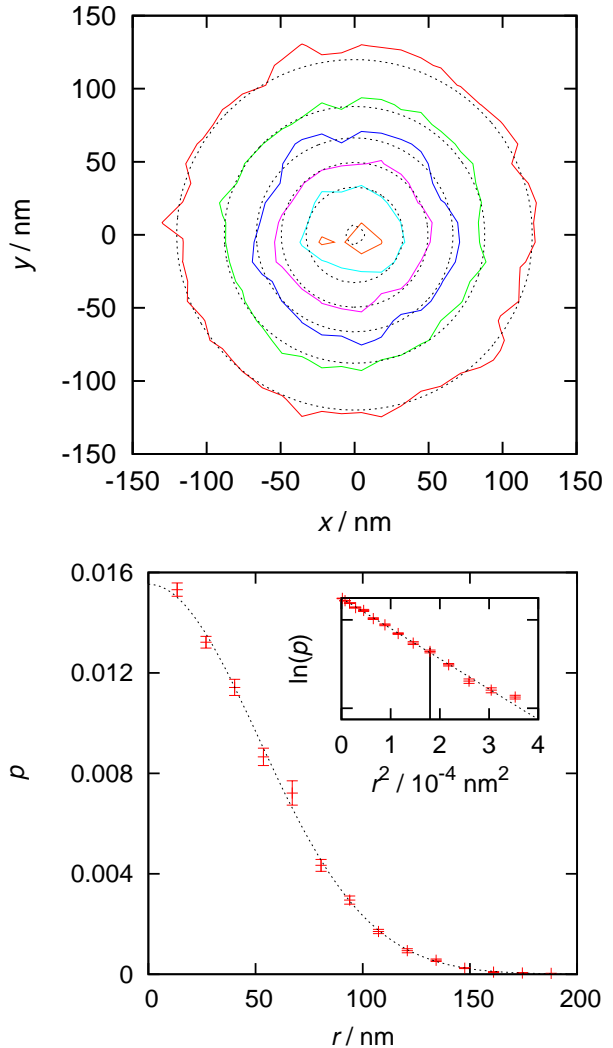}
\caption{(Colour online) \label{result:histogram} Top: Two-dimensional probability distribution of particle positions $p(r)$ with $r=(x,y)$ within the trap. Contours of the data and fit at 5\%, 20\%, 40\%, 60\%, 80\% and 99\% of maximum (outside to centre) are shown. Bottom: Azimuthally averaged probability distribution $p(|r|)$ together with a gaussian fit with variance $\sigma^{2}=(2.26 \pm 0.04) \times 10 ^{-15}$~m$^{2}$, indicating a trap stiffness $\kappa$ = $(1.56 \pm 0.03) \times 10^{-3}$~pN/nm. The inset shows the same data and fit in a log - square representation, and the vertical line shows that $\Delta x_{\mathrm{MAX}} = 134$~nm as used above is within the harmonic regime (see text for details). The same sample and experimental conditions were used as for fig~\ref{result:triangular}.}
\end{figure}

A particle or several particles were held in stationary traps and their positions determined using video microscopy and particle tracking in two dimensions. We record the position at 150~Hz for 300 s. This provides the probability distribution of the positions, $p(r)$, whose bin size was chosen to coincide with the accuracy of the particle tracking technique (fig~\ref{result:histogram}). A gaussian was fitted to $p(r)$ and from the variance $\sigma^{2}$ the trap stiffness $\mathrm{\kappa =k_{B}}T/\sigma^{2}$ determined ~\cite{Simmons1996}. With the same particle and conditions as above we determined $\sigma^{2}=(2.26 \pm 0.04) \times 10 ^{-15}$~m$^{2}$ and thus $\kappa$ = $(1.56 \pm 0.03) \times 10^{-3}$~pN/nm, which agrees within experimental uncertainty with the value determined using the active method.

\subsection{Three dimensional particle arrays}\label{sec:time}
\begin{figure}
\centering
\includegraphics[width=1.0\linewidth]{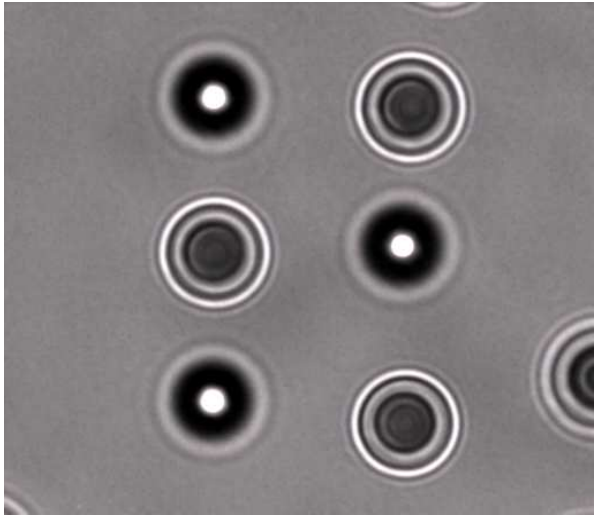}
\caption{\label{image:particles}Image of six polystyrene sulfonate particles with radius $R=2.0~\mathrm{\mu}$m, three trapped between the objective and the focal plane (e.g. top left) and three the other side of the focal plane (e.g. top right).}
\end{figure}

With our optimised SLM we can generate multiple traps (fig~\ref{image:particles}). The SLM allows us to trap particles in three dimensions. This represents an advantage over acousto-optic deflectors, with which it is possible to trap particles in more than one plane~\cite{Vossen2004}, but full control in three dimensions is not yet possible. Our set-up furthermore provides improved flexibility in moving multiple particles due to the combination of the SLM with the GMM. This combination allows for fast and smooth movements of the trap array. The time taken for one galvanometer to move through 0.5$^{\circ}$ (equal to 10.5~$\mathrm{\mu}$m in-plane translation) and settle to within 90\% of its final position is 0.65~ms (based on manufacturer's specifications). With the SLM such a movement is not possible smoothly, due to its limited refresh rate of 72~Hz (corresponding to 14~ms) for pre-calculated holograms, or, for real-time applications due to the limited processing speed of the host computer, a refresh rate of about 3~Hz (corresponding to 330~ms).

\section{Conclusions}
We have constructed optical tweezers which combine galvanometer-mounted mirrors (GMM) and a spatial light modulator (SLM). We investigated the SLM with respect to its amplitude and phase shift and optimised the tweezers set-up accordingly. The optimum incidence angle for our SLM is considerably below the typically used 45$^{\circ}$. With an optimised set-up we can create complex arrays of traps in three dimensions and move the array in a smooth and fast way using the GMM. The traps can furthermore be used for force measurements. They were calibrated using two methods, a drag force technique and a probability distribution method, which yielded consistent trap strengths (within 3\%).

\section*{Acknowledgements}
We thank J\"org Bewerunge for help with the alignment and optimisation, Beate Moser for preparing the drawings and the group of Prof. M. Padgett (Glasgow) for sharing code to control the tweezers. This work was funded by the Deutsche Forschungsgemeinshaft (DFG) within the Sonderforschungsbereich-Transregio 6, project C7, and the priority program SPP1296 `Heterogeneous Nucleation and Microstructure Formation'. RDLH acknowledges the Marie-Curie Early Stage Training Network on `Biomimetic Systems' (Contract MEST-CT-2004-504465) and the International Helmholtz Research School (IHRS) `BioSoft'.


\begin{thebibliography}{10}

\bibitem{Svoboda1994}
Svoboda, K., Block, S.~M.
Annu Rev Bioph Biomo Struct {\bf 23}, 247 (1994).

\bibitem{Sheetz1998}
Sheetz, M.~P.
\textit{Laser Tweezers in Cell Biology (Methods in Cell Biology).} Academic Press, 1998.

\bibitem{Greulich1999}
Greulich, K.~O.,
\textit{Micromanipulation by Light in Biology and Medicine.} Springer, 1999. 

\bibitem{Molloy2002}
Molloy, J.~E. and Padgett, M.~J.
Contemp Phys {\bf 43}, 241 (2002).

\bibitem{Ashkin1997}
Ashkin, A.
Proc Natl Acad Sci {\bf 94}, 4853 (1997).

\bibitem{Ashkin1970}
Ashkin, A.
Phys Rev Lett {\bf 24}, 156 (1970).

\bibitem{Ashkin1986}
Ashkin, A., Dziedzic, J.~M., Bjorkholm, J.~E., and Chu, S.
Opt Lett {\bf 11}, 288 (1986).

\bibitem{Grier2003}
Grier, D.~G.
Nature {\bf 424}, 810 (2003).

\bibitem{Otto2008}
Otto, O., Gutsche, C., Kremer, F., and Keyser, U.~F.
Rev Sci Instrum {\bf 79}, 023710 (2008).

\bibitem{Bartlett2002}
Bartlett, P. and Henderson, S.
J Phys: Condens Matter {\bf 14}, 7757 (2002).

\bibitem{Simmons1996}
Simmons, R., Finer, J., Chu, S., and Spudich, J.
Biophys J {\bf 70}, 1813 (1996).

\bibitem{Vossen2004}
Vossen, D. L.~J., van der Horst, A., Dogterom, M., and van Blaaderen, A.
Rev Sci Instrum {\bf 75}, 2960 (2004).

\bibitem{Biancaniello2006}
Biancaniello, P.~L. and Crocker, J.~C.
Rev Sci Instrum {\bf 77}, 113702 (2006).

\bibitem{Dufresne1998}
Dufresne, E.~R. and Grier, D.~G.
Rev Sci Instrum {\bf 69}, 1974 (1998).

\bibitem{Reicherter1999}
Reicherter, M., Haist, T., Wagemann, E.~U., and Tiziani, H.~J.
Opt Lett {\bf 24}, 608 (1999).

\bibitem{Liesener2000}
Liesener, J., Reicherter, M., Haist, T., and Tiziani, H.~J.
Opt Commun {\bf 185}, 77 (2000).

\bibitem{Dufresne2001}
Dufresne, E.~R., Spalding, G.~C., Dearing, M.~T., Sheets, S.~A., and Grier,
  D.~G.
Rev Sci Instrum {\bf 72}, 1810 (2001).

\bibitem{Curtis2002}
Curtis, J.~E., Koss, B.~A., and Grier, D.~G.
Opt Commun {\bf 207}, 169 (2002).

\bibitem{Sinclair2004a}
Sinclair, G., Jordan, P., Courtial, J., Padgett, M., Cooper, J., and Laczik, Z.
Opt Express {\bf 12}, 5475 (2004).

\bibitem{Neuman2004}
Neuman, K.~C. and Block, S.~M.
Rev Sci Instrum {\bf 75}, 2787 (2004).

\bibitem{Gutsche2007}
Gutsche, C., Keyser, U.~F., Kegler, K., Kremer, F., and Linse, P.
Phys Rev E {\bf 76}, 031403 (2007).

\bibitem{Ghislain1994}
Ghislain, L.~P., Switz, N.~A., and Webb, W.~W.
Rev Sci Instrum {\bf 65}, 2762 (1994).

\bibitem{Rohrbach2005a}
Rohrbach, A.
Opt Express {\bf 13}, 9695 (2005).

\bibitem{DiLeonardo2007}
Di~Leonardo, R., Ianni, F., and Ruocco, G.
Opt Express {\bf 15}, 1913 (2007).

\bibitem{Michelson1887}
Michelson, A.~A. and Morley, E.~W.
Am J Sci {\bf 34}, 333 (1887).

\bibitem{Zehnder1891}
Zehnder, L.
Z Instrumentenkd {\bf 11}, 275 (1891).

\bibitem{Mach1892}
Mach, L.
Z Instrumentenkd {\bf 12}, 89 (1892).

\bibitem{Martin-Badosa1997}
Martin-Badosa, E., Carnicer, A., Juvells, I., and Vallmitjana, S.
Meas Sci Technol {\bf 8}, 764 (1997).

\bibitem{Kuo1993}
Kuo, S. and Sheetz, M.
Science {\bf 260}, 232 (1993).

\end{thebibliography}
\end{document}